\documentclass[9pt,sigplan,screen]{acmart}

\usepackage{tikz}
\usepackage[pages=some]{background}

\usepackage{pifont}
\newcommand{\cmark}{\ding{51}}
\newcommand{\xmark}{\ding{55}}

\usepackage{array}
\newcolumntype{C}[1]{>{\centering\arraybackslash}p{#1}}
\newcolumntype{L}[1]{>{\raggedright\arraybackslash}p{#1}}
\usepackage{tikz}
\usepackage{pgfplots}

\pgfplotsset{compat=1.18}
\usepackage[table]{xcolor}
\usepackage{booktabs}
\usepackage{pifont}
\usepackage{graphicx}
\usepackage{multirow}
\usepackage{graphicx}

\usepackage{listings}
\usepackage{xcolor}

\definecolor{codegray}{RGB}{245,245,245}
\definecolor{codeborder}{RGB}{220,220,220}

\lstdefinestyle{verilogstyle}{
    language=Verilog,
    basicstyle=\ttfamily\footnotesize,
    backgroundcolor=\color{codegray},
    frame=single,
    rulecolor=\color{codeborder},
    numbers=none,
    breaklines=true,
    columns=fullflexible,
    keepspaces=true,
    showstringspaces=false,
    tabsize=4,
    xleftmargin=0.5em,
    xrightmargin=0.5em,
    framexleftmargin=0.5em,
    framexrightmargin=0.5em,
    aboveskip=0.8em,
    belowskip=0.8em
}

\definecolor{syn0}{HTML}{FF5722}
\definecolor{syn1}{HTML}{FF8A22}
\definecolor{syn2}{HTML}{FFC247}
\definecolor{syn3}{HTML}{9CFF45}
\definecolor{syn4}{HTML}{00C853}
\definecolor{syn5}{HTML}{00A83B}

\newcommand{\Syn}[1]{%
  \ifcase#1\cellcolor{syn0}\textbf{0}%
  \or\cellcolor{syn1}\textbf{1}%
  \or\cellcolor{syn2}\textbf{2}%
  \or\cellcolor{syn3}\textbf{3}%
  \or\cellcolor{syn4}\textbf{4}%
  \or\cellcolor{syn5}\textbf{5}%
  \fi
}

\renewcommand{\cmark}{\ding{51}}
\renewcommand{\xmark}{\ding{55}}
\newcommand{\na}{-}
\usepgfplotslibrary{groupplots}

\copyrightyear{2026}
\acmYear{2026}
\setcopyright{cc}
\setcctype{by}
\acmConference[MLCAD '26]{2026 ACM/IEEE International Symposium on Machine Learning for CAD}{September 07--09, 2026}{Jeju Island, Republic of Korea}
\acmBooktitle{2026 ACM/IEEE International Symposium on Machine Learning for CAD (MLCAD '26), September 07--09, 2026, Jeju Island, Republic of Korea}
\acmDOI{10.1145/3831599.3840306}
\acmISBN{979-8-4007-2878-5/2026/09}

\begin{document}

\title{ChipVerilog: A Large-Scale OpenCores-Derived Benchmark for LLM-Based Verilog RTL Generation}

\author{Yan Tan}
\email{ytan910@connect.hkust-gz.edu.cn}
\affiliation{%
Microelectronics Thrust, The Hong Kong University of Science and Technology (Guangzhou)
\country{China}
}
\author{Jiping Du}
\email{jdu428@connect.hkust-gz.edu.cn}
\affiliation{%
Microelectronics Thrust, The Hong Kong University of Science and Technology (Guangzhou)
\country{China}
}
\author{Xiangchen Meng}
\email{xmeng027@connect.hkust-gz.edu.cn}
\affiliation{%
Microelectronics Thrust, The Hong Kong University of Science and Technology (Guangzhou)
\country{China}
}
\author{Yangdi Lyu}
\email{yangdilyu@hkust-gz.edu.cn} 
\authornote{Corresponding author.}
\affiliation{%
Microelectronics Thrust, The Hong Kong University of Science and Technology (Guangzhou)
\country{China}
}

\begin{abstract}
Large language models have shown strong potential for Verilog RTL generation. However, many existing benchmarks are built from short, self-contained module-level tasks. These tasks are useful for controlled evaluation, but they do not fully capture the code scale, hierarchy, and module interactions found in practical IP and processor-core RTL. We present ChipVerilog, a description-to-Verilog generation benchmark built from OpenCores IP/core designs. The benchmark contains 64 generation targets from five design families: OR1200, double-precision FPU, MIPS-16, I2C, and CORDIC. It includes both single-module targets and cross-module targets that instantiate or interact with other RTL modules. Several targets exceed 1,000 lines of Verilog, making ChipVerilog substantially larger and structurally more complex than typical module-level suites. Each benchmark instance is constructed from a pair of specification documents and reference RTL. We extract the target functionality, write a detailed natural-language description, and manually review the description for correctness and clarity. Generated RTL is checked by compilation and validated through equivalence checking for local modules, or by simulation for integrated IP/core targets. Results show that large-scale RTL remains challenging, especially for hierarchical and cross-module designs.
\end{abstract}


\maketitle
\backgroundsetup{opacity=1, scale=1, angle=0, contents={
\begin{tikzpicture}[remember picture, overlay]
\node[anchor=north east, inner xsep=50pt, inner ysep=10pt] at (current page.north east) {
\href{https://www.acm.org/publications/policies/artifact-review-and-badging-current}{
\includegraphics[width=50pt]{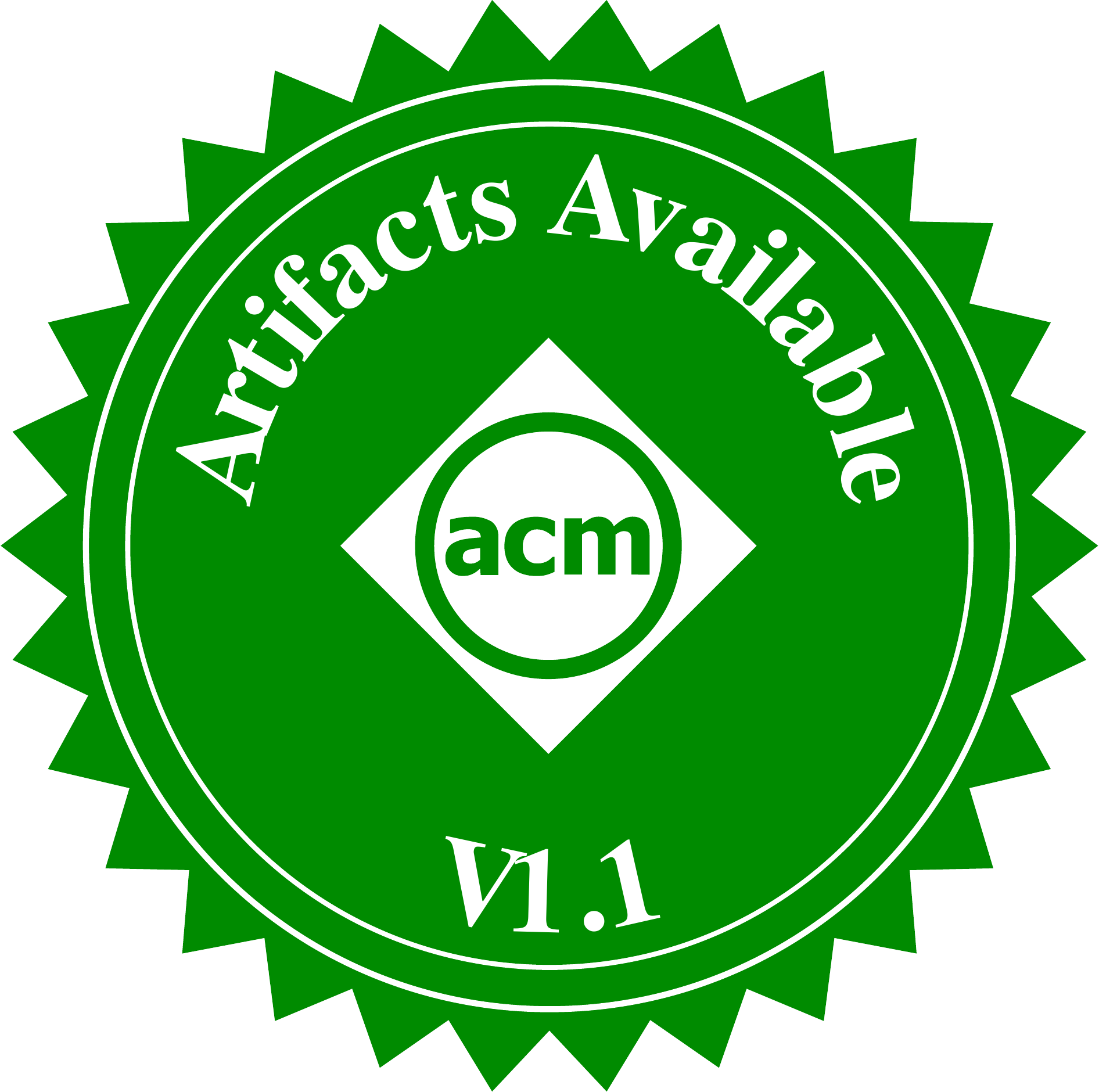}
\includegraphics[width=50pt]{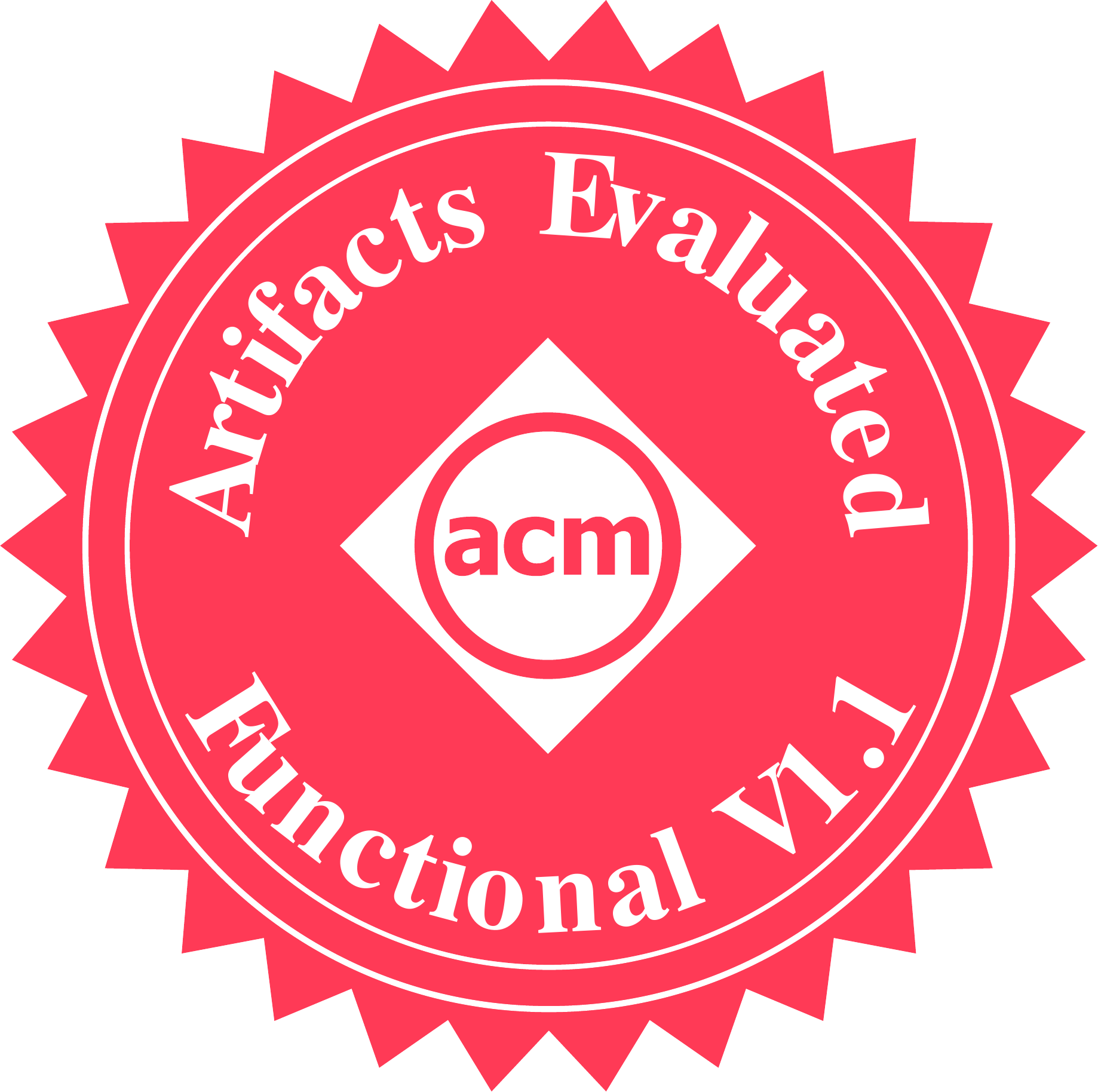}
}};
\end{tikzpicture}
}}
\BgThispage

\section{Introduction}

Recent advances in large language models (LLMs) have significantly accelerated research in hardware design automation, with promising results in Verilog RTL generation~\cite{VerilogEval, BetterV}, code optimization~\cite{2024origen, Xie_2023, RTLFixer}, debugging~\cite{mage, wang2025hlsdebugger}, and verification assistance~\cite{li2026formalrtl, bai2025assertionforge, yan2025assertllm}. These developments suggest that LLMs have the potential to become useful assistants in digital design workflows, where engineers traditionally translate specifications into RTL, refine implementations, and validate correctness under complex architectural and verification constraints. Crucially, the development and refinement of these models depend heavily on specialized hardware benchmarks, which serve as the foundational training data and fine-tuning corpora for teaching LLMs how to generate hardware designs.

However, current LLM-based Verilog benchmarks remain largely centered on simple, single-module tasks~\cite{Benchmark_RTL_2022, Dehaerne_Verilog_2023, OpenLLM, VerilogEval}. Most existing benchmark instances consist of short, self-contained RTL blocks often under 100 lines of code. Consequently, while models optimized on these datasets excel at generating isolated and flat code blocks, they fail to master the complex, deeply nested structural hierarchies inherent in modern production-grade chip designs. Furthermore, these benchmarks are typically evaluated through finite testbench simulations. While scalable, simulation-based evaluation only exercises behaviors covered by specific test stimuli, inherently leaving critical corner cases unverified. This creates a substantial gap relative to large-scale chip designs, where hardware modules are tightly integrated into broader intellectual property (IP) cores, interact via shared control paths, complex bus protocols, and memory hierarchies, and typically require formal verification guarantees.

To bridge this gap, we introduce ChipVerilog, a large-scale Verilog RTL dataset and benchmark designed for training and evaluating LLMs. ChipVerilog is constructed from manually curated OpenCores~\cite{opencores} IP and core designs, chosen for their stability, mature open-source RTL design status, and comprehensive specification documents. Each retained module is paired with a concise natural-language description derived from official documents and manually cleaned for consistency. Unlike existing datasets, ChipVerilog covers three levels of RTL targets: standalone modules, multi-module targets, and full-system targets. A standalone module is self-contained and does not instantiate other RTL modules. A multi-module target instantiates or interacts with external RTL modules, but its verification boundary is suitable for equivalence checking. A full-system target is an IP/core that integrates major subsystems such as processor pipelines, memory interfaces, bus logic, or protocol controllers. This multi-level benchmark forces LLMs to reason beyond simple code syntax, evaluating their capacity to preserve cross-module interfaces, coordinate functional dependencies, and complex integration logic.

The current version of ChipVerilog encompasses a diverse array of design domains, including arithmetic datapaths, control-intensive logic, protocol-driven interfaces, and subsystem-level integration. It contains 64 generation targets from five representative RTL design groups: a double-precision FPU, an OR1200 processor-core subsystem, a MIPS-16 processor core, an I2C controller, and a CORDIC computational core. In total, the benchmark includes approximately 20,400 lines of Verilog RTL, with a median file length of about 268 lines; 40 files exceed 200 lines, 7 exceed 500 lines, and 3 exceed 1000 lines. This scale and multi-layered structure distinguish ChipVerilog from existing simple RTL blocks, shifting the evaluation paradigm toward realistic design settings characterized by long code contexts, cross-module dependencies, and verification-driven constraints.

To summarize, our main contributions are as follows:
\begin{itemize}
\item We identify the limitations of current LLM-based RTL benchmarks, including their focus on short standalone modules, limited IP/core-level context, and heavy reliance on finite testbench simulation.

\item We introduce \textbf{ChipVerilog}\footnote{https://github.com/HKUSTGZ-MICS-LYU/ChipVerilog}, a large-scale and hierarchy-aware Verilog RTL benchmark derived from manually curated open-source IP cores, covering standalone modules, multi-module targets, and full-system targets.

\item We propose a level-aware evaluation flow that uses equivalence checking for standalone modules and multi-module targets, and simulation-based validation for full-system targets.

\item We evaluate state-of-the-art LLMs on ChipVerilog and analyze their limitations in long-context, hierarchy-aware, and integration-aware RTL generation.
\end{itemize}

\begin{table*}[ht]
\centering
\small
\setlength{\tabcolsep}{3.2pt}
\renewcommand{\arraystretch}{0.88}
\caption{Comparison of open-source description-to-Verilog benchmarks.}
\label{tab:desc2verilog_open_benchmarks}
\resizebox{\textwidth}{!}{
\begin{tabular}{lcccccc}
\toprule
Benchmark & Size & $>$1000 LOC & RTL Level & Module Interaction & Verification & Source \\
\midrule
ChipGPT~\cite{Chang_Wang_Ren_Wang_Liang_Han_Li_Li} 
& 8 & No & Module & None & Simulation & Hand-crafted \\
VeriGen\_test~\cite{Benchmark_RTL_2022} 
& 17 & No & Module & None & Simulation & Textbook / HDLBits \\
VerilogEval~\cite{VerilogEval} 
& 156 & No & Module & None & Simulation & HDLBits \\
AutoChip~\cite{AutoChip} 
& 120 & No & Module & None & Simulation & HDLBits \\
ChipGPTV~\cite{chipgptv} 
& 30 & No & Module & None & Simulation & Hand-crafted \\
Evaluate\_LLMs~\cite{blocklove2024evaluating} 
& 8 & No & Module & None & Simulation & Textbook / Hand-crafted \\
RTLLM-v1~\cite{lu2024rtllm} 
& 30 & No & Module & None & Simulation & Academic / Hand-crafted \\
RTLLM-v2~\cite{OpenLLM} 
& 50 & No & Module & None & Simulation & Academic / Hand-crafted \\
ArchXBench~\cite{purini2025archxbench} 
& 51 & No & Module / Partial IP & None & Simulation & Academic \\
CVDP~\cite{pinckney2025comprehensive} 
& 141$^\dagger$ & No & Module / Partial IP & None & Test harness & Engineer-authored \\
RealBench~\cite{jin2025realbench} 
& 60 & No & IP & Cross-module & Simulation + Formal & Open-source IPs \\
ChipVerilog (ours) 
& 64 & Yes & IP/core & Cross-module & Equivalence + Simulation & OpenCores~\cite{opencores} \\
\bottomrule
\end{tabular}
}

\vspace{1mm}

\begin{flushleft}
\footnotesize
$^\dagger$ The reported CVDP size counts only its specification-to-RTL generation tasks; other task families in CVDP, such as debugging, verification, and Q\&A, are excluded. 
\end{flushleft}

\end{table*}

\section{Related Work}

A key challenge in LLM-based Verilog RTL generation is the lack of realistic benchmarks that reflect the scale, hierarchy, and integration constraints of practical hardware designs. Compared with software code, high-quality Verilog corpora are more limited in size, diversity, and contextual richness. Many existing datasets are built from short, self-contained modules with simplified specifications, making them convenient for controlled evaluation but less representative of real RTL development, where modules often interact through shared control paths, bus protocols, memory interfaces, and hierarchical design structures.

Existing description-to-Verilog benchmarks have made steady progress. Early works such as ChipGPT~\cite{Chang_Wang_Ren_Wang_Liang_Han_Li_Li} and VeriGen~\cite{Benchmark_RTL_2022} introduced natural-language-to-Verilog generation tasks based on small module-level designs, often from textbooks or hand-crafted examples. VerilogEval~\cite{VerilogEval} and AutoChip~\cite{AutoChip} further standardized the evaluation of executables using HDLBits-style problems and testbench-based validation. Later benchmarks expanded the task scope: ChipGPTV~\cite{chipgptv} considers multimodal specifications, Evaluate\_LLMs~\cite{blocklove2024evaluating} provides a compact textbook and hand-crafted suite, and RTLLM-v1/v2~\cite{lu2024rtllm, OpenLLM} organize hand-crafted RTL tasks across multiple functional categories. More recent efforts move toward higher design realism. ArchXBench~\cite{purini2025archxbench} targets architecture-oriented RTL synthesis, CVDP~\cite{pinckney2025comprehensive} covers generation, debugging, verification, and Q\&A with structured test harnesses, and RealBench~\cite{jin2025realbench} evaluates IP-level Verilog generation using open-source IPs with both simulation and formal validation.

As summarized in Table~\ref{tab:desc2verilog_open_benchmarks}, existing benchmarks are still dominated by short, module-level tasks. Recent IP-level benchmarks improve design realism, but large RTL targets with explicit hierarchy and cross-module dependencies remain limited. \textbf{ChipVerilog} addresses this gap by using OpenCores-derived IP/core designs with longer targets, including modules over 1000 lines and designs that instantiate or depend on other RTL components.

\section{ChipVerilog}
We introduce ChipVerilog, an OpenCores-derived benchmark for evaluating LLM-based description-to-Verilog generation on large-scale IP/core RTL. This section presents ChipVerilog in four parts: an overview of the benchmark workflow, the selected RTL designs, the construction of design specifications, and the validation methodology.

\begin{figure*}
    \centering
    \includegraphics[width=1\linewidth]{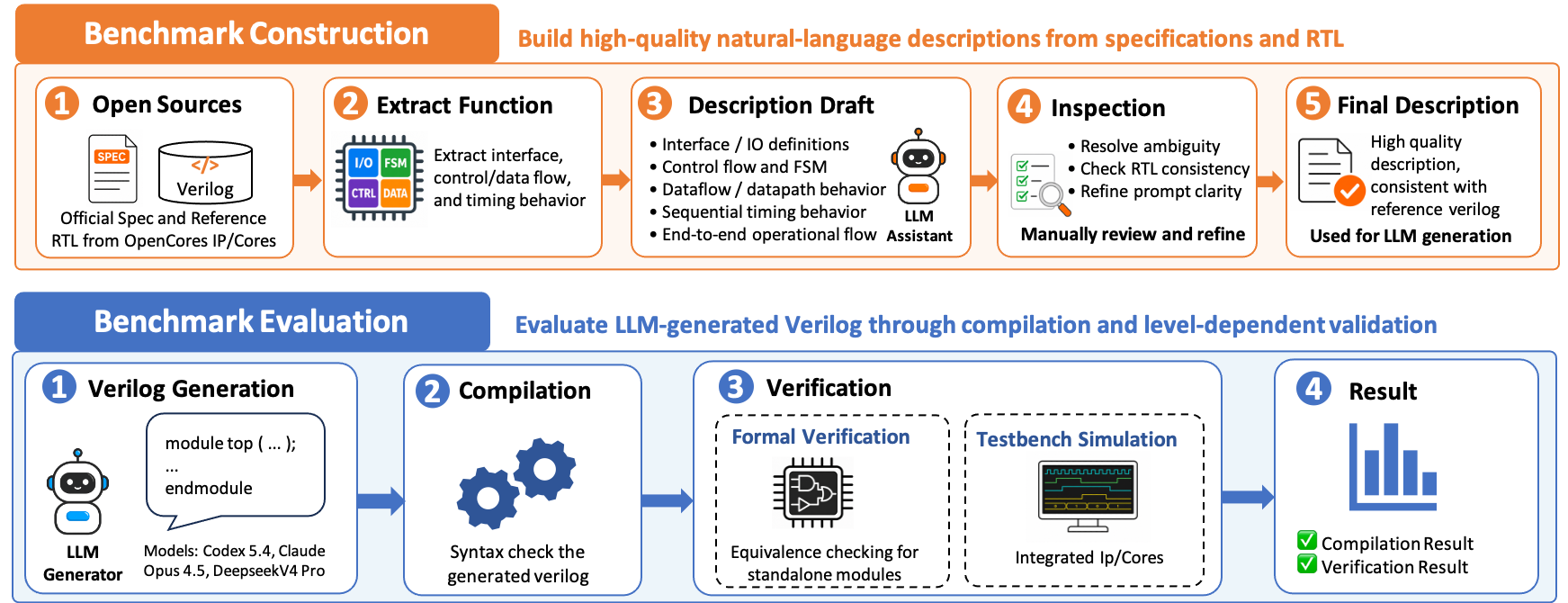}
    \caption{Overview of the ChipVerilog benchmark construction and evaluation pipeline.}
    \label{fig:workflow}
\end{figure*}

\subsection{Overview}
As illustrated in Figure~\ref{fig:workflow}, our methodology follows an end-to-end pipeline with two main phases: \textbf{Benchmark Construction} and \textbf{Benchmark Evaluation}. In the construction stage, we start from paired specification documents and reference Verilog RTL. We first extract the target module's functionality, interfaces, and key behaviors from these sources, then organize the extracted information into a detailed natural-language description. The description is manually reviewed and revised to remove ambiguity and ensure consistency with the reference RTL. The approved description is then stored as the final benchmark prompt.

In the evaluation stage, the final description is given to an LLM to generate the target Verilog RTL. The generated code first undergoes syntax, compilation, and elaboration checks. Only compilable outputs proceed to functional validation. Depending on the target type, ChipVerilog uses equivalence checking for standalone modules and multi-module targets, and simulation-based validation for full-system targets.

\subsection{Designs}

The current release contains 64 generation targets distributed across five representative RTL design groups. Macro definition files, testbenches, and purely auxiliary files are excluded from the target set.

\begin{itemize}
    \item \textbf{Floating-Point Unit (FPU):} 9 targets covering double-precision arithmetic, IEEE-754 rounding, exception handling, and pipelined floating-point operations.
    \item \textbf{OR1200 Processor Core:} 38 targets covering processor pipeline logic, cache and MMU components, exception handling, debug support, timer and interrupt logic, and Wishbone bus interfaces.
    \item \textbf{MIPS-16 Core:} 11 targets covering a compact five-stage processor pipeline, including instruction fetch, decode, execution, memory access, write-back, register-file access, and hazard detection.
    \item \textbf{I2C Controller:} 3 targets covering the top-level I2C master, byte-level control, and bit-level serial protocol control.
    \item \textbf{CORDIC Core:} 3 targets covering iterative arithmetic logic, signed shifting, and CORDIC rotation-stage computation.
\end{itemize}

These design groups cover complementary RTL styles. The FPU and CORDIC subsets emphasize arithmetic datapath generation; the MIPS-16 subset captures compact pipeline organization; the I2C subset represents protocol-control logic; and the OR1200 subset provides the richest IP/core-level hierarchy, including integration-facing modules that instantiate or coordinate other RTL components. This composition allows ChipVerilog to evaluate description-to-Verilog generation across both standalone modules and modules embedded in larger processor or IP-level structures.

\begin{table*}[t]
\centering
\footnotesize
\setlength{\tabcolsep}{0pt}
\renewcommand{\arraystretch}{0.88}
\caption{ChipVerilog benchmark description. The benchmark contains 64 generation targets across five OpenCores-derived RTL design families. Macro definition files and testbenches are excluded.}
\label{tab:chipverilogsuite_description}
\resizebox{\textwidth}{!}{
\begin{tabular}{@{}c@{\hspace{0.9em}}c@{}}
\toprule
\begin{tabular}[t]{@{}l@{\hspace{0.65em}}c@{\hspace{0.65em}}c@{\hspace{0.65em}}p{0.19\textwidth}@{}}
\multicolumn{4}{c}{\textbf{FPU Modules}} \\
\cmidrule(lr){1-4}
\textbf{Design} & \textbf{Lines} & \textbf{Hier.} & \textbf{Description} \\
\midrule
fpu\_add & 130 & single & Double-precision addition \\
fpu\_div & 412 & single & Double-precision division \\
fpu\_double & 302 & multi(7) & Top-level double-precision FPU \\
fpu\_exceptions & 280 & single & IEEE-754 exception handling \\
fpu\_mul & 273 & single & Double-precision multiplication \\
fpu\_round & 91 & single & IEEE-754 rounding and packing \\
fpu\_sub & 221 & single & Double-precision subtraction \\
fpu\_addsub\_pipeline & 367 & single & Pipelined add/subtract \\
fpu\_mul\_pipeline & 270 & single & Pipelined multiplication \\

\midrule
\multicolumn{4}{c}{\textbf{MIPS-16 Modules}} \\
\cmidrule(lr){1-4}
\textbf{Design} & \textbf{Lines} & \textbf{Hier.} & \textbf{Description} \\
\midrule
EX\_stage & 59 & multi(1) & Execution pipeline stage \\
ID\_stage & 290 & single & Instruction decode stage \\
IF\_stage & 56 & multi(1) & Instruction fetch stage \\
MEM\_stage & 62 & multi(1) & Memory access stage \\
WB\_stage & 44 & single & Register write-back stage \\
alu & 51 & single & 16-bit ALU \\
data\_mem & 39 & single & Data memory \\
hazard\_detection\_unit & 57 & single & Pipeline hazard detection \\
instruction\_mem & 80 & single & Instruction memory \\
mips\_16\_core\_top & 126 & multi(7) & Top-level MIPS-16 core \\
register\_file & 81 & single & Register file \\

\midrule
\multicolumn{4}{c}{\textbf{I2C Modules}} \\
\cmidrule(lr){1-4}
\textbf{Design} & \textbf{Lines} & \textbf{Hier.} & \textbf{Description} \\
\midrule
i2c\_master\_top & 300 & multi(1) & I2C master top module \\
i2c\_master\_byte\_ctrl & 344 & multi(1) & Byte-level I2C controller \\
i2c\_master\_bit\_ctrl & 576 & single & Bit-level I2C controller \\

\midrule
\multicolumn{4}{c}{\textbf{CORDIC Modules}} \\
\cmidrule(lr){1-4}
\textbf{Design} & \textbf{Lines} & \textbf{Hier.} & \textbf{Description} \\
\midrule
cordic & 379 & multi(4) & Configurable CORDIC core \\
signed\_shifter & 179 & single & Signed shift unit \\
rotator & 253 & multi(2) & CORDIC rotation stage \\
\end{tabular}
&
\begin{tabular}[t]{@{}l@{\hspace{0.65em}}c@{\hspace{0.65em}}c@{\hspace{0.65em}}p{0.19\textwidth}@{}}
\multicolumn{4}{c}{\textbf{OR1200 Modules}} \\
\cmidrule(lr){1-4}
\textbf{Design} & \textbf{Lines} & \textbf{Hier.} & \textbf{Description} \\
\midrule
or1200\_alu & 468 & single & Execution-stage ALU \\
or1200\_cfgr & 234 & single & Configuration register interface \\
or1200\_cpu & 806 & multi(13) & Main OR1200 pipeline core \\
or1200\_ctrl & 1052 & single & Instruction decode and control \\
or1200\_dc\_fsm & 323 & single & Data-cache control FSM \\
or1200\_dc\_ram & 189 & multi(3) & Data-cache data array \\
or1200\_dc\_tag & 181 & multi(3) & Data-cache tag array \\
or1200\_dc\_top & 344 & multi(3) & Data-cache integration \\
or1200\_dmmu\_tlb & 343 & multi(4) & Data-side TLB \\
or1200\_dmmu\_top & 349 & multi(1) & Data MMU integration \\
or1200\_du & 1798 & multi(1) & Debug unit \\
or1200\_except & 579 & single & Exception/interrupt handling \\
or1200\_freeze & 203 & single & Pipeline freeze control \\
or1200\_genpc & 346 & single & Next-PC generation \\
or1200\_gmultp2\_32x32 & 130 & single & 32$\times$32 signed multiplier \\
or1200\_ic\_fsm & 265 & single & Instruction-cache control FSM \\
or1200\_ic\_ram & 196 & multi(4) & Instruction-cache data array \\
or1200\_ic\_tag & 198 & multi(4) & Instruction-cache tag array \\
or1200\_ic\_top & 353 & multi(3) & Instruction-cache integration \\
or1200\_if & 189 & single & Instruction fetch logic \\
or1200\_immu\_tlb & 352 & multi(4) & Instruction-side TLB \\
or1200\_immu\_top & 422 & multi(1) & Instruction MMU integration \\
or1200\_iwb\_biu & 516 & single & Instruction-side Wishbone BIU \\
or1200\_lsu & 197 & multi(2) & Load/store unit \\
or1200\_mem2reg & 433 & single & Memory-to-register logic \\
or1200\_mult\_mac & 338 & multi(2) & Multiplier/MAC unit \\
or1200\_operandmuxes & 184 & single & Operand selection logic \\
or1200\_pic & 226 & single & Interrupt controller \\
or1200\_pm & 215 & single & Power-management unit \\
or1200\_qmem\_top & 475 & multi(1) & QMEM local-memory interface \\
or1200\_reg2mem & 134 & single & Register-to-memory alignment \\
or1200\_rf & 456 & multi(4) & Register file \\
or1200\_sb & 193 & multi(1) & Store buffer \\
or1200\_sprs & 477 & single & SPR interface \\
or1200\_top & 1065 & multi(13) & Top-level OR1200 wrapper \\
or1200\_tt & 223 & single & Tick timer \\
or1200\_wb\_biu & 478 & single & Data-side Wishbone BIU \\
or1200\_wbmux & 168 & single & Write-back mux \\
\end{tabular}
\\
\bottomrule
\end{tabular}
}
\vspace{1mm}
\begin{flushleft}
\footnotesize
\emph{Hier.} denotes module hierarchy: \emph{single} indicates a standalone target, while \emph{multi(n)} indicates a target involving $n$ instantiated or dependent submodules.
\end{flushleft}
\end{table*}

\subsection{Specification Curation and Description Design}

The quality and granularity of the input specification strongly affect the reliability of description-to-Verilog generation. Existing benchmarks often use concise functional requests, textbook-style summaries, or simplified HDLBits-style prompts. In contrast, ChipVerilog constructs benchmark-ready natural-language specifications by combining information from official specification documents and reference RTL implementations. Following Steps 2--5 in Figure~\ref{fig:workflow}, we curate each prompt through function extraction, description construction, and manual review.

Each specification is organized around interface, behavior, and execution information. The interface part defines input/output ports, signal widths, clock/reset assumptions, and macro-dependent interface variations. The behavior part describes the module's dataflow, control flow, finite-state-machine transitions, arithmetic or protocol operations, configuration-dependent behavior, and sequential timing. The execution part summarizes the end-to-end processing flow, connecting the module inputs, internal control/data paths, state updates, and output behavior.

The manual review step checks whether the specification is consistent with the reference RTL and sufficiently clear for generation. During this process, we resolve ambiguous wording, remove irrelevant implementation noise, and avoid exposing the reference implementation verbatim. The resulting prompt captures the intended functionality and key implementation constraints of the target module while still requiring the model to synthesize the corresponding Verilog RTL.

\begin{itemize}
    \item \textbf{Interface and boundary definitions:} input/output ports, signal widths, clock/reset assumptions, macro-controlled ports, and integration-facing interface constraints.
    \item \textbf{Dataflow and datapath behavior:} data dependencies, datapath transformations, arithmetic operations, mux selections, register updates, and output computation rules.
    \item \textbf{Control-flow and FSM behavior:} control conditions, branch decisions, state transitions, enable/stall behavior, exception paths, and protocol-level actions.
    \item \textbf{Timing and sequential behavior:} clocked update rules, pipeline stages, handshake timing, multi-cycle dependencies, and reset or hold behavior.
    \item \textbf{Operational flow:} a step-by-step description that summarizes how the module processes inputs, updates internal state, coordinates control/data paths, and produces outputs.
\end{itemize}

\subsection{Validation Flow}
\label{sec:validation}
During evaluation, the finalized description is fed to an LLM to generate the target Verilog RTL. The generated code undergoes initial syntax validation, compilation, and elaboration. Designs that fail this stage are immediately classified as syntax failures and excluded from functional verification. For successfully compiled designs, ChipVerilog employs a level-aware validation strategy tailored to structural complexity, categorized into three distinct tiers.

For standalone modules, we apply formal equivalence checking against the reference RTL. This establishes a more rigorous correctness criterion than finite testbench simulation alone. For multi-module targets, only the target module is replaced with the generated RTL, while the remaining submodules are kept as reference implementations for equivalence checking. This verifies the generated module's interface behavior and interactions with surrounding RTL components. For full-system targets, the target is an IP/core-level entry point that integrates major subsystems, such as processor pipelines, memory interfaces, or protocol controllers. Comprehensive formal equivalence checking becomes impractical for these targets because of the large state spaces. Therefore, full-system targets are evaluated through simulation-based validation using testbenches.

\begin{table*}[t]
\centering
\scriptsize
\setlength{\tabcolsep}{3pt}
\renewcommand{\arraystretch}{0.88}
\caption{Per-instance syntax and functional results. Syntax results report the number of compilable samples out of five. Functional results indicate whether at least one generated sample passes functional validation.}
\label{tab:benchmark_results_heatmap}
\resizebox{\textwidth}{!}{
\begin{tabular}{l|cc|cc|cc|l|cc|cc|cc}
\toprule
\multirow{2}{*}[-0.3ex]{\footnotesize\textbf{Design}}
& \multicolumn{2}{c|}{\footnotesize\textbf{Claude}}
& \multicolumn{2}{c|}{\footnotesize\textbf{GPT-5.4}}
& \multicolumn{2}{c|}{\footnotesize\textbf{DeepSeek}}
& \multirow{2}{*}[-0.3ex]{\footnotesize\textbf{Design}}
& \multicolumn{2}{c|}{\footnotesize\textbf{Claude}}
& \multicolumn{2}{c|}{\footnotesize\textbf{GPT-5.4}}
& \multicolumn{2}{c|}{\footnotesize\textbf{DeepSeek}}\\
\cmidrule(lr){2-3} \cmidrule(lr){4-5} \cmidrule(lr){6-7}
\cmidrule(lr){9-10} \cmidrule(lr){11-12} \cmidrule(lr){13-14}
& Syn. & Func. & Syn. & Func. & Syn. & Func.
& & Syn. & Func. & Syn. & Func. & Syn. & Func. \\
\midrule
or1200\_alu & \Syn{4} & \cmark & \Syn{3} & \xmark & \Syn{1} & \xmark & or1200\_reg2mem & \Syn{5} & \cmark & \Syn{5} & \cmark & \Syn{4} & \xmark \\
or1200\_cfgr & \Syn{5} & \xmark & \Syn{3} & \xmark & \Syn{0} & \na & or1200\_rf & \Syn{5} & \xmark & \Syn{4} & \xmark & \Syn{0} & \na \\
or1200\_cpu & \Syn{4} & \xmark & \Syn{5} & \xmark & \Syn{0} & \na & or1200\_sb & \Syn{5} & \cmark & \Syn{5} & \cmark & \Syn{3} & \cmark \\
or1200\_ctrl & \Syn{5} & \xmark & \Syn{4} & \xmark & \Syn{0} & \na & or1200\_sprs & \Syn{4} & \xmark & \Syn{3} & \xmark & \Syn{3} & \xmark \\
or1200\_dc\_fsm & \Syn{5} & \xmark & \Syn{5} & \xmark & \Syn{4} & \xmark & or1200\_top & \Syn{5} & \xmark & \Syn{5} & \xmark & \Syn{0} & \na \\
or1200\_dc\_ram & \Syn{5} & \xmark & \Syn{5} & \xmark & \Syn{3} & \xmark & or1200\_tt & \Syn{3} & \cmark & \Syn{4} & \cmark & \Syn{1} & \xmark \\
or1200\_dc\_tag & \Syn{5} & \xmark & \Syn{5} & \xmark & \Syn{3} & \xmark & fpu\_mul & \Syn{5} & \xmark & \Syn{5} & \xmark & \Syn{1} & \xmark \\
or1200\_dc\_top & \Syn{3} & \xmark & \Syn{2} & \xmark & \Syn{0} & \na & fpu\_sub & \Syn{4} & \xmark & \Syn{5} & \cmark & \Syn{2} & \cmark \\
or1200\_dmmu\_tlb & \Syn{4} & \xmark & \Syn{5} & \xmark & \Syn{2} & \xmark & fpu\_add & \Syn{4} & \xmark & \Syn{5} & \cmark & \Syn{5} & \xmark \\
or1200\_dmmu\_top & \Syn{3} & \xmark & \Syn{3} & \xmark & \Syn{0} & \na & fpu\_div & \Syn{4} & \xmark & \Syn{5} & \xmark & \Syn{1} & \xmark \\
or1200\_du & \Syn{3} & \xmark & \Syn{2} & \xmark & \Syn{1} & \xmark & fpu\_double & \Syn{0} & \na & \Syn{0} & \na & \Syn{0} & \na \\
or1200\_except & \Syn{5} & \xmark & \Syn{4} & \xmark & \Syn{4} & \xmark & fpu\_exceptions & \Syn{5} & \cmark & \Syn{5} & \xmark & \Syn{2} & \xmark \\
or1200\_freeze & \Syn{5} & \cmark & \Syn{5} & \cmark & \Syn{5} & \cmark & fpu\_round & \Syn{5} & \cmark & \Syn{5} & \cmark & \Syn{4} & \xmark \\
or1200\_genpc & \Syn{2} & \xmark & \Syn{3} & \xmark & \Syn{4} & \xmark & fpu\_addsub\_pipeline & \Syn{5} & \cmark & \Syn{5} & \cmark & \Syn{3} & \xmark \\
or1200\_gmultp2\_32x32 & \Syn{5} & \cmark & \Syn{5} & \cmark & \Syn{3} & \cmark & fpu\_mul\_pipeline & \Syn{4} & \xmark & \Syn{5} & \xmark & \Syn{0} & \na \\
or1200\_ic\_fsm & \Syn{5} & \cmark & \Syn{5} & \cmark & \Syn{3} & \cmark & EX\_stage & \Syn{4} & \cmark & \Syn{5} & \cmark & \Syn{0} & \na \\
or1200\_ic\_ram & \Syn{4} & \xmark & \Syn{3} & \xmark & \Syn{1} & \xmark & ID\_stage & \Syn{3} & \xmark & \Syn{5} & \xmark & \Syn{4} & \xmark \\
or1200\_ic\_tag & \Syn{5} & \xmark & \Syn{5} & \xmark & \Syn{4} & \xmark & IF\_stage & \Syn{1} & \cmark & \Syn{5} & \cmark & \Syn{2} & \cmark \\
or1200\_ic\_top & \Syn{3} & \xmark & \Syn{5} & \xmark & \Syn{0} & \na & MEM\_stage & \Syn{4} & \cmark & \Syn{5} & \xmark & \Syn{3} & \cmark \\
or1200\_if & \Syn{5} & \cmark & \Syn{5} & \cmark & \Syn{5} & \xmark & WB\_stage & \Syn{4} & \cmark & \Syn{5} & \cmark & \Syn{5} & \cmark \\
or1200\_immu\_tlb & \Syn{4} & \xmark & \Syn{5} & \xmark & \Syn{1} & \xmark & alu & \Syn{4} & \cmark & \Syn{4} & \cmark & \Syn{3} & \xmark \\
or1200\_wb\_biu & \Syn{4} & \xmark & \Syn{5} & \xmark & \Syn{4} & \xmark & data\_mem & \Syn{5} & \cmark & \Syn{5} & \cmark & \Syn{5} & \cmark \\
or1200\_wbmux & \Syn{5} & \xmark & \Syn{5} & \cmark & \Syn{3} & \cmark & hazard\_detection\_unit & \Syn{4} & \cmark & \Syn{5} & \cmark & \Syn{5} & \cmark \\
or1200\_immu\_top & \Syn{2} & \xmark & \Syn{3} & \xmark & \Syn{3} & \xmark & instruction\_mem & \Syn{2} & \xmark & \Syn{4} & \xmark & \Syn{5} & \xmark \\
or1200\_iwb\_biu & \Syn{3} & \xmark & \Syn{5} & \xmark & \Syn{1} & \xmark & register\_file & \Syn{4} & \cmark & \Syn{5} & \cmark & \Syn{4} & \cmark \\
or1200\_lsu & \Syn{5} & \xmark & \Syn{4} & \xmark & \Syn{0} & \na & mips\_16\_core\_top & \Syn{2} & \xmark & \Syn{0} & \na & \Syn{0} & \na \\
or1200\_mem2reg & \Syn{5} & \xmark & \Syn{3} & \cmark & \Syn{3} & \xmark & i2c\_master\_top & \Syn{3} & \xmark & \Syn{5} & \xmark & \Syn{1} & \xmark \\
or1200\_mult\_mac & \Syn{4} & \xmark & \Syn{3} & \xmark & \Syn{0} & \na & i2c\_master\_byte\_ctrl & \Syn{4} & \cmark & \Syn{5} & \xmark & \Syn{2} & \xmark \\
or1200\_operandmuxes & \Syn{3} & \cmark & \Syn{4} & \cmark & \Syn{3} & \xmark & i2c\_master\_bit\_ctrl & \Syn{5} & \xmark & \Syn{5} & \xmark & \Syn{3} & \xmark \\
or1200\_pic & \Syn{4} & \cmark & \Syn{4} & \cmark & \Syn{2} & \cmark & rotator & \Syn{1} & \xmark & \Syn{0} & \na & \Syn{1} & \xmark \\
or1200\_pm & \Syn{5} & \cmark & \Syn{5} & \cmark & \Syn{4} & \cmark & signed\_shifter & \Syn{0} & \na & \Syn{0} & \na & \Syn{4} & \xmark \\
or1200\_qmem\_top & \Syn{5} & \cmark & \Syn{4} & \cmark & \Syn{1} & \cmark & cordic & \Syn{4} & \xmark & \Syn{5} & \xmark & \Syn{3} & \xmark \\
\bottomrule
\end{tabular}
}
\end{table*}

\section{Experimental Results}

\subsection{Experimental Setup}

We evaluate ChipVerilog using generation targets collected from OpenCores~\cite{opencores} IP/core RTL designs. Each benchmark instance consists of a curated natural-language specification paired with a target Verilog module. Macro definition files, testbenches, and auxiliary support files are excluded from the generation target set. The benchmark includes standalone modules, multi-module targets, and full-system targets. We use Icarus Verilog~\cite{icarus_verilog} for syntax checking, compilation, elaboration, and simulation, and Yosys~\cite{wolf2013yosys} for equivalence checking. We evaluate three state-of-the-art LLMs: Claude Opus 4.5~\cite{anthropic_claude_opus_45_2025}, GPT-5.4~\cite{openai_gpt54_2026}, and DeepSeek V4 Pro~\cite{deepseekai2026deepseekv4}. Generated RTL is first checked using Icarus Verilog (\texttt{iverilog}) for syntax validity, compilation, and elaboration. 

Following the validation strategy in Section~\ref{sec:validation}, standalone modules and multi-module targets are evaluated using Yosys-based equivalence checking, where generated RTL replaces only the target module and required submodules are kept from the reference design. Full-system targets are evaluated using \texttt{iverilog}-based testbench simulation. The full-system targets are \texttt{or1200\_top}, \texttt{mips\_16\_core\_top}, \texttt{i2c\_master\_top}, \texttt{fpu\_double}, and \texttt{cordic}; all remaining targets are evaluated by equivalence checking.

Following VerilogEval~\cite{VerilogEval}, we use the metric \textit{pass@k} to evaluate functional correctness. \textit{pass@k} represents the expected probability that at least one generated sample passes functional validation when randomly selecting $k$ samples from $n$ generated candidates. We follow the standard unbiased estimator used in VerilogEval and set $n=5$ in all experiments. 

\subsection{Overall Results}

\begin{table}[t]
\centering
\setlength{\tabcolsep}{6pt}
\renewcommand{\arraystretch}{0.88}
\caption{Average syntax and functional pass@k results.}
\label{tab:pass_at_k}
\begin{tabular}{l|cc|cc}
\toprule
Model 
& \multicolumn{2}{c|}{Syntax} 
& \multicolumn{2}{c}{Function} \\
\cmidrule(lr){2-3} \cmidrule(lr){4-5}
& pass@1 & pass@5 & pass@1 & pass@5 \\
\midrule
Claude Opus 4.5~\cite{anthropic_claude_opus_45_2025} & 78.75\% & 96.88\% & 17.50\% & 35.94\% \\
GPT-5.4~\cite{openai_gpt54_2026} & 83.13\% & 93.75\% & 23.13\% & 37.50\% \\
DeepSeek V4 Pro~\cite{deepseekai2026deepseekv4} & 45.94\% & 78.13\% & 13.44\% & 23.44\% \\
\bottomrule
\end{tabular}
\end{table}

\begin{figure*}[t]
\centering

\definecolor{syntaxBlue}{RGB}{70,105,230}
\definecolor{syntaxFill}{RGB}{190,205,255}
\definecolor{funcRed}{RGB}{230,95,80}
\definecolor{funcFill}{RGB}{255,205,195}

\resizebox{\textwidth}{!}{%
\begin{tikzpicture}
\begin{groupplot}[
    group style={
        group size=4 by 1,
        horizontal sep=0.30cm
    },
    ybar,
/pgf/bar width=6pt,
    width=0.30\textwidth,
    height=4.8cm,
    ymin=0,
    ymax=105,
    symbolic x coords={<100,100--300,300--500,500+},
    xtick=data,
    xticklabel style={
        rotate=25,
        anchor=east,
        font=\scriptsize
    },
    xlabel={Code Length},
    xlabel style={font=\small, yshift=1pt},
    ylabel style={font=\small},
    tick label style={font=\scriptsize},
    title style={font=\small\bfseries},
    enlarge x limits=0.16,
    axis line style={black!55},
    ytick style={draw=none},
    xtick style={draw=none},
    grid=none,
    nodes near coords,
    nodes near coords align={vertical},
    every node near coord/.append style={
        font=\scriptsize,
        /pgf/number format/fixed,
        /pgf/number format/precision=1
    },
    legend columns=2,
    legend style={
        draw=none,
        font=\small,
        column sep=1.0em
    },
]

\nextgroupplot[
    title={(a) Overall},
    ylabel={Pass Rate (\%)},
    legend to name=passlegend,
]
\addplot[
    draw=syntaxBlue,
    fill=syntaxFill,
    line width=0.7pt,
    nodes near coords style={text=syntaxBlue}
] coordinates {
    (<100,80.7) (100--300,71.8) (300--500,61.6) (500+,65.7)
};

\addplot[
    draw=funcRed,
    fill=funcFill,
    line width=0.7pt,
    nodes near coords style={text=funcRed}
] coordinates {
    (<100,58.0) (100--300,17.7) (300--500,3.2) (500+,0.0)
};

\legend{Syntax Pass@1, Functional Pass@1}

\nextgroupplot[
    title={(b) Claude Opus 4.5 ~\cite{anthropic_claude_opus_45_2025}},
    ylabel={},
    yticklabels={},
]
\addplot[
    draw=syntaxBlue,
    fill=syntaxFill,
    line width=0.7pt,
    nodes near coords style={text=syntaxBlue}
] coordinates {
    (<100,74.0) (100--300,82.3) (300--500,73.3) (500+,85.7)
};

\addplot[
    draw=funcRed,
    fill=funcFill,
    line width=0.7pt,
    nodes near coords style={text=funcRed}
] coordinates {
    (<100,62.0) (100--300,15.4) (300--500,3.8) (500+,0.0)
};

\nextgroupplot[
    title={(c) GPT 5.4~\cite{openai_gpt54_2026}},
    ylabel={},
    yticklabels={},
]
\addplot[
    draw=syntaxBlue,
    fill=syntaxFill,
    line width=0.7pt,
    nodes near coords style={text=syntaxBlue}
] coordinates {
    (<100,96.0) (100--300,82.3) (300--500,77.1) (500+,85.7)
};

\addplot[
    draw=funcRed,
    fill=funcFill,
    line width=0.7pt,
    nodes near coords style={text=funcRed}
] coordinates {
    (<100,68.0) (100--300,22.3) (300--500,4.8) (500+,0.0)
};

\nextgroupplot[
    title={(d) DeepSeek V4 Pro~\cite{deepseekai2026deepseekv4}},
    ylabel={},
    yticklabels={},
]
\addplot[
    draw=syntaxBlue,
    fill=syntaxFill,
    line width=0.7pt,
    nodes near coords style={text=syntaxBlue}
] coordinates {
    (<100,72.0) (100--300,50.8) (300--500,34.3) (500+,25.7)
};

\addplot[
    draw=funcRed,
    fill=funcFill,
    line width=0.7pt,
    nodes near coords style={text=funcRed}
] coordinates {
    (<100,44.0) (100--300,15.3) (300--500,0.9) (500+,0.0)
};

\end{groupplot}
\end{tikzpicture}
}
\vspace{1mm}
\pgfplotslegendfromname{passlegend}

\caption{Syntax pass@1 and functional pass@1 grouped by reference RTL length.}
\label{fig:length_pass_all_models}
\end{figure*}
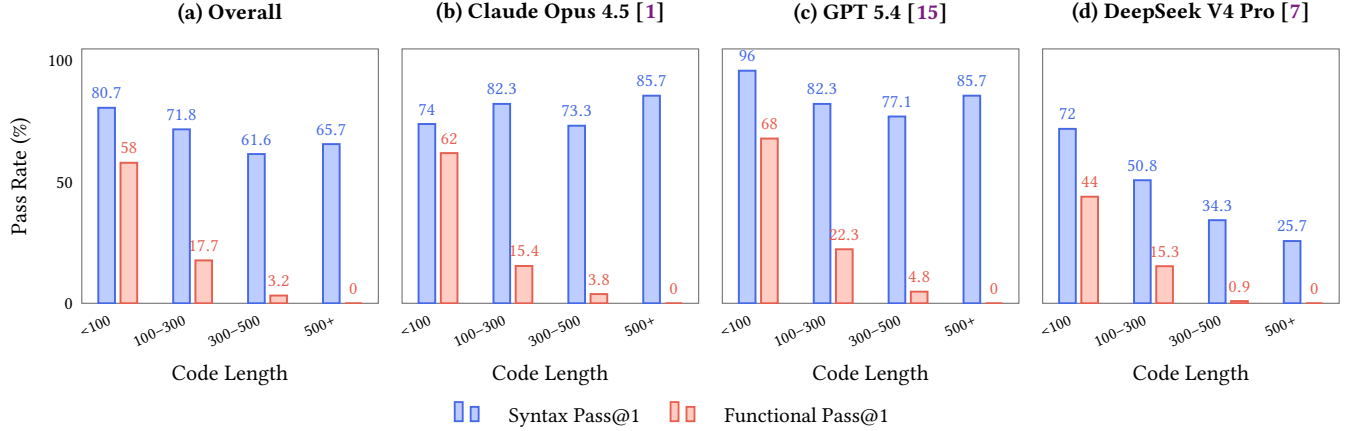

Tables~\ref{tab:benchmark_results_heatmap} and~\ref{tab:pass_at_k} summarize the overall results. Across all models, syntax pass rates are much higher than functional pass rates, showing that current LLMs can often generate compilable Verilog but still fail to preserve the reference RTL behavior.

GPT-5.4 achieves the best pass@1 results, with 83.13\% syntax pass@1 and 23.13\% functional pass@1. Claude Opus 4.5 follows with 78.75\% syntax pass@1 and 17.50\% functional pass@1, while DeepSeek V4 Pro obtains 45.94\% and 13.44\%, respectively. Increasing the number of samples improves all models, especially
on syntax: Claude reaches 96.88\% syntax pass@5, GPT-5.4 reaches 93.75\%, and DeepSeek reaches 78.13\%. Functional pass@5 remains substantially lower, reaching 35.94\% for Claude, 37.50\% for GPT-5.4, and 23.44\% for DeepSeek. This gap shows that additional sampling helps recover syntax-level failures but still does not reliably produce behaviorally correct RTL.

This gap between syntax and functional correctness is central to our findings. Multiple samples recover many syntax-level failures, but do not reliably produce behaviorally correct RTL. Claude and GPT-5.4 both exceed 93\% syntax pass@5, while their functional pass@5 values are 35.94\% and 37.50\%, respectively. The
per-instance results further show that failures are more frequent on long and hierarchy-rich targets, including multi-module and full-system targets. These results indicate that ChipVerilog exposes challenges beyond compilability, requiring models to preserve control behavior, interface constraints, timing assumptions, and cross-module interactions.

\begin{table}
\centering
\caption{Pass@1 and pass@5 grouped by the number of instantiated or dependent submodules. 
The results are aggregated across all evaluated models.}
\renewcommand{\arraystretch}{0.88}
\label{tab:submodule_passk}
\begin{tabular}{c|c|cc|cc}
\toprule
Submodules & Tasks 
& \multicolumn{2}{c|}{Syntax} 
& \multicolumn{2}{c}{Function} \\
\cmidrule(lr){3-4} \cmidrule(lr){5-6}
 & & pass@1 & pass@5 & pass@1 & pass@5 \\
\midrule
0  & 37 & 77.48\% & 95.50\% & 23.06\% & 44.14\% \\
1  & 10 & 61.33\% & 93.33\% & 25.33\% & 46.67\% \\
2+ & 17 & 55.69\% & 74.51\% &  0.00\% &  0.00\% \\
\bottomrule
\end{tabular}
\end{table}

\subsection{Impact of Code Length}

Figure~\ref{fig:length_pass_all_models} groups the results by reference RTL length. Syntax pass@1 remains relatively high across length groups, ranging from 61.6\% to 80.7\%, while functional pass@1 drops sharply as code length increases: 58.0\% for targets below 100 lines, 17.7\% for 100--300 lines, 3.2\% for 300--500 lines, and 0.0\% for targets above 500 lines. These results indicate that increasing code length mainly challenges semantic correctness rather than syntactic generation. Current LLMs can still produce compilable Verilog for many long modules, but they often fail to preserve detailed control behavior, timing dependencies, interface constraints, and corner cases. Claude Opus 4.5 and GPT-5.4 maintain relatively strong syntax pass rates across most length groups, while DeepSeek V4 Pro is more sensitive to longer code. However, all three models exhibit near-zero functional correctness on long targets, highlighting the difficulty of long-context RTL generation.

\subsection{Impact of Module Hierarchy}

Table~\ref{tab:submodule_passk} reports hierarchy-level results aggregated across all evaluated models. Standalone targets and one-submodule targets achieve similar syntax pass@5 rates, 95.50\% and 93.33\%, respectively, with functional pass@5 rates of 44.14\% and 46.67\%. The main degradation appears for targets with two or more submodules. Although their syntax pass@5 remains 74.51\%, both functional pass@1 and pass@5 are 0.00\%. These results indicate that deeper hierarchy introduces challenges beyond compilability, including port wiring, width matching, timing alignment, and cross-module control/data coordination.

\section{Conclusion}

We presented \textbf{ChipVerilog}, an OpenCores-derived benchmark for evaluating LLM-based description-to-Verilog generation on large-scale IP/core RTL. Unlike short standalone-module benchmarks, ChipVerilog includes long RTL targets across standalone modules, multi-module targets, and full-system targets, covering floating-point units, processor cores, bus-control logic, and iterative arithmetic modules. Our results show a clear gap between syntactic validity and functional correctness. Although strong models achieve high syntax pass rates, their functional rates remain much lower, especially for long modules and targets involving multiple submodules. This indicates that current LLMs can often generate compilable Verilog, but still struggle with control behavior, interface consistency, and integration-level RTL semantics. ChipVerilog complements existing benchmarks by exposing these challenges in realistic OpenCores IP/core settings. Future work will expand the benchmark to more IP families and incorporate more detailed failure diagnosis and feedback-based repair evaluation.

\bibliographystyle{ACM-Reference-Format}
\bibliography{references}

@article{Dehaerne_Verilog_2023,
  title={A deep learning framework for verilog autocompletion towards design and verification automation},
  author={Dehaerne, Enrique and Dey, Bappaditya and Halder, Sandip and De Gendt, Stefan},
  journal={arXiv preprint arXiv:2304.13840},
  year={2023}
}

@article{BetterV,
  title={BetterV: Controlled verilog generation with discriminative guidance},
  author={Pei, Zehua and Zhen, Hui-Ling and Yuan, Mingxuan and Huang, Yu and Yu, Bei},
  journal={arXiv preprint arXiv:2402.03375},
  year={2024}
}

@inproceedings{Benchmark_RTL_2022,
  title={Benchmarking large language models for automated verilog rtl code generation
  },
  author={Thakur, Shailja and Ahmad, Baleegh and Fan, Zhenxing and Pearce, Hammond and Tan, Benjamin and Karri, Ramesh and Dolan-Gavitt, Brendan and Garg, Siddharth},
  booktitle={2023 Design, Automation \& Test in Europe Conference \& Exhibition (DATE)},
  pages={1--6},
  year={2023},
  organization={IEEE}
}

@inproceedings{VerilogEval,
  title={VerilogEval: Evaluating large language models for verilog code generation},
  author={Liu, Mingjie and Pinckney, Nathaniel and Khailany, Brucek and Ren, Haoxing},
  booktitle={2023 IEEE/ACM International Conference on Computer Aided Design (ICCAD)},
  pages={1--8},
  year={2023},
  organization={IEEE}
}

@inproceedings{OpenLLM,
  title={{OpenLLM-RTL}: Open Dataset and Benchmark for LLM-Aided Design RTL Generation(Invited)},
  author={Liu, Shang and Lu, Yao and Fang, Wenji and Li, Mengming and Xie, Zhiyao},
  booktitle={2024 IEEE/ACM International Conference on Computer-Aided Design (ICCAD)},
  year={2024},
  organization={ACM}
}

@inproceedings{lu2024rtllm,
  title={{RTLLM}: An open-source benchmark for design rtl generation with large language model},
  author={Lu, Yao and Liu, Shang and Zhang, Qijun and Xie, Zhiyao},
  booktitle={2024 29th Asia and South Pacific Design Automation Conference (ASP-DAC)},
  pages={722--727},
  year={2024},
  organization={IEEE}
}

@article{wang2025hlsdebugger,
  title={{HLSDebugger}: Identification and Correction of Logic Bugs in HLS Code with LLM Solutions},
  author={Wang, Jing and Liu, Shang and Lu, Yao and Xie, Zhiyao},
  journal={arXiv preprint arXiv:2507.21485},
  year={2025}
}

@inproceedings{Xie_2023,
  title={{RtlCoder}: Outperforming gpt-3.5 in design rtl generation with our open-source dataset and lightweight solution},
  author={Liu, Shang and Fang, Wenji and Lu, Yao and Zhang, Qijun and Zhang, Hongce and Xie, Zhiyao},
  booktitle={2024 IEEE LLM Aided Design Workshop (LAD)},
  pages={1--5},
  year={2024},
  organization={IEEE}
}

@article{AutoChip,
  title={Autochip: Automating hdl generation using llm feedback},
  author={Thakur, Shailja and Blocklove, Jason and Pearce, Hammond and Tan, Benjamin and Garg, Siddharth and Karri, Ramesh},
  journal={arXiv preprint arXiv:2311.04887},
  year={2023}
}

@inproceedings{RTLFixer,
  title={{RTLFixer}: Automatically fixing RTL syntax errors with large language model},
  author={Tsai, YunDa and Liu, Mingjie and Ren, Haoxing},
  booktitle={Proceedings of the 61st ACM/IEEE Design Automation Conference},
  pages={1--6},
  year={2024}
}

@article{2024origen,
  title={{OriGen}: Enhancing rtl code generation with code-to-code augmentation and self-reflection},
  author={Cui, Fan and Yin, Chenyang and Zhou, Kexing and Xiao, Youwei and Sun, Guangyu and Xu, Qiang and Guo, Qipeng and Song, Demin and Lin, Dahua and Zhang, Xingcheng and others},
  journal={arXiv preprint arXiv:2407.16237},
  year={2024}
}

@article{Chang_Wang_Ren_Wang_Liang_Han_Li_Li,
  title={Chipgpt: How far are we from natural language hardware design},
  author={Chang, Kaiyan and Wang, Ying and Ren, Haimeng and Wang, Mengdi and Liang, Shengwen and Han, Yinhe and Li, Huawei and Li, Xiaowei},
  journal={arXiv preprint arXiv:2305.14019},
  year={2023}
}

@INPROCEEDINGS{mage,
  author={Zhao, Yujie and Zhang, Hejia and Huang, Hanxian and Yu, Zhongming and Zhao, Jishen},
  booktitle={2025 62nd ACM/IEEE Design Automation Conference (DAC)}, 
  title={MAGE: A Multi-Agent Engine for Automated RTL Code Generation}, 
  year={2025},
  volume={},
  number={},
  pages={1-7},
  doi={10.1109/DAC63849.2025.11133191}}

@inproceedings{chipgptv,
  title={Natural language is not enough: Benchmarking multi-modal generative AI for Verilog generation},
  author={Chang, Kaiyan and Chen, Zhirong and Zhou, Yunhao and Zhu, Wenlong and Wang, Kun and Xu, Haobo and Li, Cangyuan and Wang, Mengdi and Liang, Shengwen and Li, Huawei and others},
  booktitle={Proceedings of the 43rd IEEE/ACM International Conference on Computer-Aided Design},
  pages={1--9},
  year={2024}
}

@inproceedings{blocklove2024evaluating,
  title={Evaluating llms for hardware design and test},
  author={Blocklove, Jason and Garg, Siddharth and Karri, Ramesh and Pearce, Hammond},
  booktitle={2024 IEEE LLM Aided Design Workshop (LAD)},
  pages={1--6},
  year={2024},
  organization={IEEE}
}

@inproceedings{purini2025archxbench,
  title={ArchXBench: A Complex Digital Systems Benchmark Suite for LLM Driven RTL Synthesis},
  author={Purini, Suresh and Garg, Siddhant and Gaur, Mudit and Bhat, Sankalp and Mupparapu, Sohan and Ravindran, Arun},
  booktitle={2025 ACM/IEEE 7th Symposium on Machine Learning for CAD (MLCAD)},
  pages={1--10},
  year={2025},
  organization={IEEE}
}

@article{pinckney2025comprehensive,
  title={Comprehensive Verilog design problems: A next-generation benchmark dataset for evaluating large language models and agents on rtl design and verification},
  author={Pinckney, Nathaniel and Deng, Chenhui and Ho, Chia-Tung and Tsai, Yun-Da and Liu, Mingjie and Zhou, Wenfei and Khailany, Brucek and Ren, Haoxing},
  journal={arXiv preprint arXiv:2506.14074},
  year={2025}
}

@article{jin2025realbench,
  title={Realbench: Benchmarking verilog generation models with real-world ip designs},
  author={Jin, Pengwei and Huang, Di and Li, Chongxiao and Cheng, Shuyao and Zhao, Yang and Zheng, Xinyao and Zhu, Jiaguo and Xing, Shuyi and Dou, Bohan and Zhang, Rui and others},
  journal={arXiv preprint arXiv:2507.16200},
  year={2025}
}

@article{li2026formalrtl,
  title={FormalRTL: Verified RTL Synthesis at Scale},
  author={Li, Kezhi and Li, Min and Wen, Xiangyu and Zhao, Shibo and Wu, Jieying and Huang, Junhua and Xu, Qiang},
  journal={arXiv preprint arXiv:2603.08738},
  year={2026}
}

@inproceedings{bai2025assertionforge,
  title={Assertionforge: Enhancing formal verification assertion generation with structured representation of specifications and rtl},
  author={Bai, Yunsheng and Hamad, Ghaith Bany and Suhaib, Syed and Ren, Haoxing},
  booktitle={2025 IEEE International Conference on LLM-Aided Design (ICLAD)},
  pages={85--92},
  year={2025},
  organization={IEEE}
}

@inproceedings{yan2025assertllm,
  title={Assertllm: Generating hardware verification assertions from design specifications via multi-llms},
  author={Yan, Zhiyuan and Fang, Wenji and Li, Mengming and Li, Min and Liu, Shang and Xie, Zhiyao and Zhang, Hongce},
  booktitle={Proceedings of the 30th Asia and South Pacific Design Automation Conference},
  pages={614--621},
  year={2025}
}

@misc{opencores,
  author       = {{OpenCores}},
  title        = {{OpenCores}: Open Source Hardware IP Cores},
  year         = {2026},
  url          = {https://opencores.org/},
  note         = {Accessed: 2026-05}
}

@misc{anthropic_claude_opus_45_2025,
  author       = {{Anthropic}},
  title        = {{Claude Opus 4.5}},
  year         = {2025},
  month        = nov,
  howpublished = {\url{https://www.anthropic.com/claude/opus}},
  note         = {Accessed: 2026-05-20}
}

@misc{openai_gpt54_2026,
  author       = {{OpenAI}},
  title        = {{GPT-5.4 Model}},
  year         = {2026},
  howpublished = {\url{https://developers.openai.com/api/docs/models/gpt-5.4}},
  note         = {Accessed: 2026-05-20}
}

@misc{deepseekai2026deepseekv4,
      title={DeepSeek-V4: Towards Highly Efficient Million-Token Context Intelligence},
      author={DeepSeek-AI},
      year={2026},
}

@misc{icarus_verilog,
  author       = {Williams, Stephen},
  title        = {{Icarus Verilog}},
  howpublished = {\url{https://steveicarus.github.io/iverilog/}},
  note         = {Accessed: 2026-05-20}
}

@inproceedings{wolf2013yosys,
  author    = {Wolf, Clifford and Glaser, Johann and Kepler, Johannes},
  title     = {{Yosys -- A Free Verilog Synthesis Suite}},
  booktitle = {Proceedings of the 21st Austrian Workshop on Microelectronics},
  year      = {2013}
}

\clearpage
\appendix

\section{Artifact Appendix}
\label{sec:artifact_appendix}

\subsection{Artifact Overview}

The ChipVerilog artifact accompanies the benchmark and evaluation presented
in this paper. It provides the benchmark specifications, reference RTL,
generated RTL candidates, and the verification resources used to evaluate
description-to-Verilog generation on realistic IP/core designs.

The artifact is organized to support both inspection of individual benchmark
tasks and reproduction of the reported evaluation. In particular, each task
pairs a curated natural-language specification with its corresponding RTL
target and associated verification context.

\subsection{Artifact Organization}

Each benchmark instance is organized around three main components:

\begin{itemize}
    \item \textbf{Specification:}
    a curated natural-language description of the target module, including its
    interface, functional behavior, timing requirements, and relevant
    integration constraints.

    \item \textbf{Reference design:}
    the corresponding Verilog RTL implementation used as the functional
    reference during evaluation.

    \item \textbf{Evaluation context:}
    the supporting RTL modules, testbenches, or verification scripts required
    to validate the generated implementation under the appropriate benchmark
    boundary.
\end{itemize}

This organization preserves the distinction between standalone modules,
hierarchical modules that depend on surrounding RTL, and full-system targets
whose behavior is validated at the integration level.

\subsection{Representative Task Structure}

A benchmark prompt is designed to describe enough implementation-relevant
behavior for RTL generation without exposing the reference implementation
verbatim. A typical task specification contains the following information:

\begin{itemize}
    \item \textbf{Module objective:}
    the role of the target module within the corresponding IP/core.

    \item \textbf{Interface definition:}
    input and output ports, signal widths, clock/reset assumptions, and
    interface-level constraints.

    \item \textbf{Functional behavior:}
    datapath operations, control decisions, state transitions, and output
    behavior.

    \item \textbf{Operation encodings:}
    command, mode, or control-field meanings when the target module exposes
    encoded operations.

    \item \textbf{Submodule interaction:}
    the external RTL components instantiated or coordinated by the target
    module, together with the role of each dependency.

    \item \textbf{Timing behavior:}
    sequential update rules, pipeline behavior, fixed- or variable-latency
    conditions, and completion signaling.

    \item \textbf{Processing flow:}
    an end-to-end description connecting inputs, internal control/data paths,
    intermediate results, and final outputs.
\end{itemize}

For example, the representative floating-point task described in
Appendix~A specifies the top-level objective, the operand and control
interfaces, operation and rounding encodings, arithmetic submodule routing,
rounding and exception handling, and operation-dependent completion latency.
This illustrates the level of behavioral and integration detail used
throughout the benchmark.

\subsection{Reproducibility Notes}

The released artifact contains the exact benchmark prompts and the RTL
candidates used for the evaluation reported in this paper. Cached generated
outputs are retained so that the reported results can be inspected and
re-evaluated without requiring access to hosted language-model services.

The verification setup follows the evaluation methodology described in the
main paper. Standalone and locally bounded hierarchical targets are checked
against their reference behavior, while full-system targets are evaluated
within their corresponding integration test environments.

The artifact therefore separates model generation from result validation:
reviewers and users may directly inspect or re-evaluate the archived RTL
without regenerating model outputs.

\subsection{Artifact Availability}

The evaluated artifact snapshot is archived on Zenodo: \url{https://doi.org/10.5281/zenodo.21946068}. The project repository is available at: \url{https://github.com/HKUSTGZ-MICS-LYU/ChipVerilog}

\end{document}